\documentclass[11pt,a4paper]{article}
\usepackage{jheppub}
\usepackage{graphicx}
\usepackage{epstopdf}
\usepackage{amsmath}
\usepackage{amsfonts}
\usepackage{amssymb}
\usepackage{appendix}
\usepackage{comment}
\usepackage{bbold}
\usepackage{xcolor}
\definecolor{paramcol}{HTML}{006600}
\definecolor{avgcol}{HTML}{2697AD}
\definecolor{shapecol}{HTML}{CC2C0E}
\definecolor{pertcol}{HTML}{660066}
\definecolor{firstcol}{HTML}{D60ED6}
\definecolor{dpcol}{HTML}{C96840}
\usepackage{slashed}
\usepackage{subfigure}
\usepackage{setspace}
\usepackage{footnote}
\usepackage{multirow}
\usepackage{longtable}
\usepackage{braket}

\begin{document}

\singlespacing

{\hfill NUHEP-TH/18-02, FERMILAB-PUB-18-019-T\\ \today}

\title{Matter Density Profile Shape Effects at DUNE}

\author[a]{Kevin J. Kelly,}
\author[b]{Stephen J. Parke}
\affiliation[a]{Northwestern University, Department of Physics \& Astronomy, 2145 Sheridan Road, Evanston, IL 60208, USA}
\affiliation[b]{Theoretical Physics Department, Fermi National Accelerator Laboratory, P.O. Box 500, Batavia, IL 60510, USA}
\emailAdd{kjk@u.northwestern.edu}
\emailAdd{parke@fnal.gov}

\abstract{Quantum mechanical interactions between neutrinos and matter along the path of propagation, the Wolfenstein matter effect, are of particular importance for the upcoming long-baseline neutrino oscillation experiments, specifically the Deep Underground Neutrino Experiment (DUNE). Here, we explore specifically what about the matter density profile can be measured by DUNE, considering both the shape and normalization of the profile between the neutrinos' origin and detection. Additionally, we explore the capability of a perturbative method for calculating neutrino oscillation probabilities and whether this method is suitable for DUNE. We also briefly quantitatively explore the ability of DUNE to measure the Earth's matter density, and the impact of performing this measurement on measuring standard neutrino oscillation parameters.}


\maketitle

\setcounter{equation}{0}
\section{Introduction}
\label{sec:Introduction}
Neutrino physics is entering a new era of precision measurements, following up on the discovery that neutrinos have mass and leptons mix. Neutrino oscillations are a particularly interesting direction by which one can study physics beyond that predicted by the Standard Model (SM) of particle physics. One experiment that will carry the field into this new era is the Deep Underground Neutrino Experiment (DUNE)~\cite{Acciarri:2015uup}, which expects to begin collecting data within the next decade. DUNE is one of several next-generation long-baseline neutrino experiments that has been proposed to continue the quest to precisely measure neutrino oscillations.

The experimental goals of DUNE include measuring the neutrino mass ordering, the octant of the atmospheric mixing angle (whether the third neutrino mass eigenstate is composed more of muon- or tau-flavor neutrino), and whether there is CP violation in the lepton sector through the phase $\delta$. Existing experiments have begun to make progress towards all of these goals, however there is no definitive answer yet for any, and no truly definitive answer will likely be given before DUNE begins its experimental run. Of key importance for these goals at DUNE is the fact that its long-baseline consists of matter that the neutrinos have the opportunity to interact with while travelling, a non-trivial effect that impacts neutrino oscillations in a measurable way. These impacts have been well-studied for several decades~\cite{Wolfenstein:1977ue} and are critical for the physics goals of the experiment (See, e.g., Refs.~\cite{Coloma:2014kca,Nath:2015kjg,Das:2016bwe,He:2016dco,DeRomeri:2016qwo,Das:2017fcz,Kolupaeva:2017nmc}). However, recent discussion has arisen over how well-known the Earth's matter density is known, and whether this uncertainty can impact the ability of DUNE to perform its experimental goals~\cite{Roe:2017zdw}.

In this paper, we address uncertainties in the Earth's matter density profile and the measurement capability of DUNE. We show that, while matter density effects are important for its experimental goals, changes to the neutrino oscillation by changing the profile in ways discussed in Ref.~\cite{Roe:2017zdw} will not be realizable at DUNE. Previous works, such as Refs.~\cite{Ohlsson:2001et,Jacobsson:2001zk,Jacobsson:2002nb,Brahmachari:2003bk,Shan:2003vh,Ohlsson:2003ip}, have explored the impact on oscillation probabilities from a changing matter density. Here, we focus specifically on the impact at DUNE. Additionally, we discuss a perturbative method for calculating neutrino oscillations, first introduced in Ref.~\cite{Denton:2016wmg}, and analyze how suitable it is for DUNE. We see that this method is simultaneously capable of calculating probabilities for the sake of DUNE and several orders of magnitude faster in calculation than conventional, more exact methods.

This manuscript is organized as follows: in Section~\ref{sec:Oscillations}, we review the framework in which neutrino oscillation probabilities are calculated, as well as how matter density effects impact these probabilities. In Section~\ref{sec:Precision}, we analyze the oscillation probability measurement precision in a number of ways -- in Section~\ref{subsec:Naive}, we perform a na{\"i}ve statistical argument for this precision. We improve on this estimate in Section~\ref{subsec:OscParams} by analyzing how well DUNE will be able to measure oscillation parameters. In Section~\ref{subsec:MatterDensity}, we explore the change of oscillation probabilities caused by changing the matter density profile's average density and shape, and in Section~\ref{subsec:PerturbativeSensitivity}, we see how precisely the perturbative method discussed can calculate oscillation probabilities. In Section~\ref{sec:Conclusions}, we offer some concluding remarks.

\setcounter{equation}{0}
\section{Neutrino Oscillations in Matter}
\label{sec:Oscillations}

Oscillations between flavor eigenstates of neutrinos occur during propagation due to the difference in masses between mass eigenstates and the sizable mismatch between the two eigenbases. We characterize this mismatch using the PMNS Matrix $U$, where $\ket{\nu_\alpha} = U_{\alpha i} \ket{\nu_i}$. Here, Greek indices $\alpha = e, \mu, \tau$ refer to the flavor basis, and Latin indices $i = 1, 2, 3$ refer to the mass basis. Where oscillations are concerned, the matrix $U$ depends on three mixing angles $\theta_{12}$, $\theta_{13}$, and $\theta_{23}$, as well as one CP-violating phase $\delta$.

The probability that a neutrino, produced in a flavor-diagonal interaction as a state $\nu_\alpha$, travels a distance $L$, and has oscillated into a state $\nu_\beta$, then, is
\begin{equation}
P_{\alpha\beta} \equiv \left\lvert \bra{\nu_\beta} U e^{-i H_{ij} L} U^\dagger \ket{\nu_\alpha} \right\rvert^2, \label{eq:Osc}
\end{equation}
where $H_{ij}$, assumed to be constant, is the Hamiltonian in the mass eigenbasis. This additionally assumes that the neutrinos travel ultrarelativistically. In vacuum, $H_{ij} \equiv 1/(2 E_\nu) \mathrm{diag} \left\lbrace 0, \Delta m_{21}^2, \Delta m_{31}^2\right\rbrace$, where $E_\nu$ is the energy of the neutrino and $\Delta m_{ji}^2 \equiv m_j^2 - m_i^2$ is the neutrino mass-squared splitting.

During propagation through Earth, interactions between neutrinos and the electrons, neutrons, and protons induce an effective interaction potential $V$, diagonal in the flavor basis. Interactions with neutrons and protons are identical for all neutrino flavors, however there is an asymmetry between interactions of $\nu_e$ with electrons compared to interactions of $\nu_{\mu,\tau}$. Because of this, we write $V_{\alpha\beta} = (a/2 E_\nu) \mathrm{diag}\left\lbrace 1, 0, 0 \right\rbrace,$ where $a = 2\sqrt{2} G_F n_e E_\nu$, and $G_F$ is the Fermi constant and $n_e$ is the number density of electrons in the path of propagation, again, assumed here to be constant. Writing $n_e$ in terms of matter density $\rho$ and electron fraction $Y_e$,
\begin{equation}
a \simeq 1.52 \times 10^{-4} \left(\frac{Y_e \rho}{\mathrm{g/cm}^3}\right) \left(\frac{E_\nu}{\mathrm{GeV}}\right) \mathrm{eV}^2.
\end{equation}
For the remainder of this work, we assume $Y_e = 1/2$. Comparing with the measured mass-squared splittings, $a$ will be on the comparable to $\Delta m_{31}^2$ for GeV-scale $E_\nu$. The propagation Hamiltonian is then modified, $H_{ij} \rightarrow H_{ij} + U_{i\alpha}^\dagger V_{\alpha\beta} U_{\beta j}$. For antineutrinos oscillating, the probability is calculated in the same way, however $U \rightarrow U^*$ and $a \rightarrow -a$.

In Eq.~(\ref{eq:Osc}), the term $e^{-i H_{ij} L}$ is the time-evolution of the initial neutrino state as it travels over a distance $L$. As stated above, this assumes that $H_{ij}$ is constant over the entire path and $t = L$. With a varying Hamiltonian, the Schr\"{o}dinger equation\footnote{We note here that the Hamiltonian in Eq.~(\ref{eq:Schrodinger}) is written in the flavor basis, or $H_{\alpha\beta} = U_{\alpha i} H_{ij} U_{j\beta}^\dagger$.} must be solved:
\begin{equation}
i \frac{\partial}{\partial x} \ket{\nu} = H\ket{\nu}, \quad \ket{\nu} = \left(\begin{array}{c} \nu_e \\ \nu_\mu \\ \nu_\tau \end{array} \right).\label{eq:Schrodinger}
\end{equation}
Instead of solving this equation for a varying $H$, we instead treat the matter potential, and therefore the Hamiltonian, as a piecewise-constant function. Then, we can apply a series of time-evolution operators to the initial state, arriving at the following oscillation probability:
\begin{equation}
P_{\alpha\beta} = \left\lvert \bra{\nu_\beta} U \left(\prod_{n =1}^N e^{-i H_{ij}^{(n)} L_n}\right) U^\dagger \ket{\nu_\alpha} \right\rvert^2,
\end{equation}
where $N$ is the number of divisions taken along the path of propagation, $H_{ij}^{(n)}$ is the Hamiltonian and $L_n$ is the length for the $n$th division, respectively. If one takes the limit $N\to \infty$, the resulting oscillation probability agrees with that from the Schr\"{o}dinger equation.


\section{Sensitivity of DUNE}
\label{sec:Precision}
In this Section, we discuss the sensitivity of the upcoming Deep Underground Neutrino Experiment~\cite{Acciarri:2015uup} (DUNE) to changes in oscillation probabilities for neutrinos travelling the $1285$ km between Fermilab and the Sanford Underground Research Facility in South Dakota. We will be interested in the capability of the experiment to measure an oscillation probability $P_{\alpha\beta}$, and the changes in the probability that will be statistically measurable. We will refer to these changes as $|\Delta P_{\alpha\beta}|$. First, we will do so using a na{\"i}ve statistical estimate, and then we will do so by considering the stated neutrino oscillation parameter precision of the experiment. After doing so, we will discuss how changes to the matter density profile along the path of propagation can lead to measurable changes in probability, both in changing the shape and average density of the profile. Finally, we will consider a perturbative approach and discuss whether it is precise enough for the sake of DUNE.

Throughout, we will use specific colors when discussing changes in probabilities induced by a certain effect: we will use \textbf{black} for our na{\"i}ve statistical estimate, \textbf{{\color{paramcol} green}} for changes of oscillation parameters, \textbf{{\color{avgcol} blue}} for changes to the matter density profile average density, \textbf{{\color{shapecol} red}} for changes to its shape, and \textbf{{\color{pertcol} purple}} for the perturbative method at zeroth-order. We will briefly discuss the perturbative method at first-order, and will do so in \textbf{{\color{firstcol} pink}}.

\subsection{Na{\"i}ve Estimate of Sensitivity}
\label{subsec:Naive}
Let us consider that a measurement of the oscillation probability for a given channel $P_{\alpha\beta}$ is being measured for neutrinos of some energy $E_\nu$. The number of events $N$ measured for this energy will be
\begin{equation}
N = N_1 \left(E_\nu; ...\right) \times P_{\alpha\beta},
\end{equation}
where $N_1$ is the number of events that would be measured if the oscillation probability is $1$. It is a product of the neutrino flux, cross-section, detection efficiencies, etc. If we assume that the only uncertainty on $N$ is statistical and $N_1$ is well-known, then $\sigma_N = \sqrt{N}$ and $\sigma_N/N = |\Delta P_{\alpha\beta}|/P_{\alpha\beta}$. We can then substitute and arrive at our desired result,
\begin{equation}
|\Delta P_{\alpha\beta}| = \sqrt{\frac{P_{\alpha\beta}}{N_1}}.\label{eq:StatPrecision}
\end{equation}
We see here that the experiment is sensitive to smaller changes $|\Delta P_{\alpha\beta}|$ when the oscillation probability itself is lower\footnote{We note here that for a small enough probability $P_{\alpha\beta} = 1/N_1$, the number of events measured is $1$: assuming only statistical uncertainty, one cannot improve on a measurement of $1$ event.}, and that in order to be sensitive to an order of magnitude lower $|\Delta P_{\alpha\beta}|$, an experiment requires a factor of $100$ larger $N_1$.

Using Ref.~\cite{Acciarri:2015uup}, we estimate that for the energy ranges of interest at DUNE, $E_\nu \simeq 1 - 4$ GeV, $N_1 \simeq 10^3$ for both appearance and disappearance channels.
\begin{figure}[!htbp]
\centering
\includegraphics[width=0.6\linewidth]{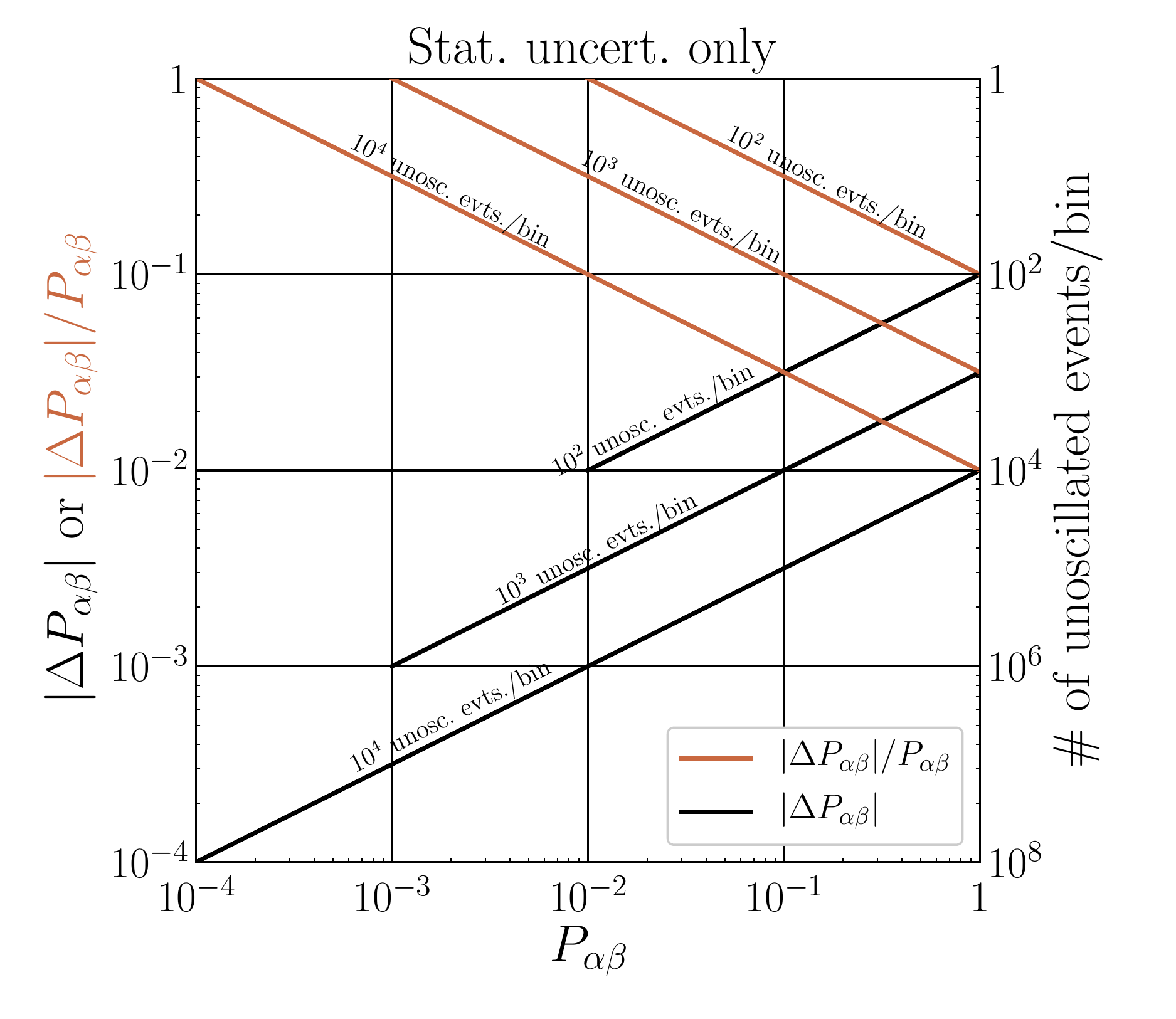}
\caption{Measurement precision of a single-bin experiment with only statistical uncertainty as a function of the oscillation probability $P_{\alpha\beta}$. \textbf{{\color{dpcol} Orange}} lines give precision in terms of fractional uncertainty $|\Delta P_{\alpha\beta}|/P_{\alpha\beta}$ where solid lines give precision in terms of $|\Delta P_{\alpha\beta}|$. The right axis, along with annotations, denotes the number of unoscillated events $N_1$ necessary in a bin to attain the given precision.}
\label{fig:PrecisionStat}
\end{figure}
In Fig.~\ref{fig:PrecisionStat}, we display the sensitivity to changes in probability $|\Delta P_{\alpha\beta}|$ for $N_1 = 10^{2}$, $10^{3}$, and $10^{4}$. Additionally, we display in \textbf{{\color{dpcol} orange}} the corresponding fractional uncertainty on the probability measurement, $|\Delta P_{\alpha\beta}|/P_{\alpha\beta}$. For the fractional uncertainty as well, two orders of magnitude larger $N_1$ is necessary to improve the sensitivity by one order of magnitude.

This process can be repeated assuming systematic uncertainties on $N_1$. We perform this exercise in Appendix~\ref{sec:SystAppendix}. The results here remain true when including this systematic uncertainty: improvement of an order of magnitude on $|\Delta P_{\alpha\beta}|$ require at least two orders of magnitude larger $N_1$ in light of this uncertainty.

DUNE will not be measuring oscillation probabilities in a single bin, but across $30$ bins in each of four channels (neutrino and antineutrino appearance and disappearance). If the measured oscillation probability $P_{\alpha\beta}$ is identical across $m$ measurements, one expects the sensitivity $|\Delta P_{\alpha\beta}|$ to decrease by a factor of $\sqrt{m}$. At DUNE, not only does the probability change across energies, the number of unoscillated events $N_1$ decreases away from the energy range of interest. We estimate this bin-to-bin measurement improvement factor to be $\sqrt{5}$ -- the measurement of the probability is being made predominantly in one bin with two bins on either side in $E_\nu$. In Fig.~\ref{fig:ParamsAndNaive}, we display the expected precision\footnote{Here, we use the oscillation parameters to be discussed cf Table~\ref{tab:Params} and calculate oscillation probabilities $P_{\mu e}$ and $P_{\mu\mu}$ as a function of neutrino energy $E_\nu$. We then use the estimated formulas for $|\Delta P_{\alpha\beta}|$ in Eqs.~(\ref{eq:StatPrecision}) and (\ref{eq:SystPrecision}) with $N_1 = 10^{3}$.} assuming $N_1 = 10^{3}$ unoscillated events, including an improvement factor of $\sqrt{5}$. We display this for appearance and disappearance channel sensitivity, both with and without a 5\% systematic uncertainty on $N_1$. We will be comparing this na{\"i}ve estimate with the sensitivity to $|\Delta P_{\alpha\beta}|$ that comes from changing oscillation parameters in Section~\ref{subsec:OscParams}. We note here that the sensitivity to the appearance channel $|P_{\mu e}|$ flattens out at $E_\nu \simeq 1.25$ GeV because $P_{\mu e} \simeq 0$ and roughly one event would be measured in this bin. In general, we expect sensitivity to $|\Delta P_{\mu e}| \simeq 3\times 10^{-3}$ (except for near $E_\nu = 1.3$ GeV, where the oscillation probability $P_{\mu e} \simeq 0$). For the disappearance channel, at all energies of interest, the sensitivity to $|\Delta P_{\mu\mu}|$ is larger than $2 \times 10^{-3}$. One could improve on these estimates with a more thorough calculation of this bin-to-bin improvement, and also by folding in the true varying $N_1$ as a function of neutrino energy.

\subsection{Sensitivity to Oscillation Parameters}
\label{subsec:OscParams}
In this subsection, we analyze changes to the neutrino oscillation probabilities that arise when parameters change, and the capability of DUNE to measure these changes. Due to the range of energies at DUNE and its baseline, the experiment will not be sensitive to the solar sector parameters $\Delta m_{21}^2$ or $\theta_{12}$. It will have significant precision in measuring the four remaining oscillation parameters; $\theta_{13}$, $\theta_{23}$, $\Delta m_{31}^2$, and $\delta$. In Table~\ref{tab:Params}, we summarize the expected precision of the experiment to measuring these four parameters, assuming the true values listed, as detailed in Ref.~\cite{Acciarri:2015uup}.
\begin{table}[!htbp]
\begin{center}
\begin{tabular}{|c||c|c|}\hline
Parameter & Physical Value & $1\sigma$ Range \\ \hline \hline
$\sin^2\theta_{23}$ & $0.450$ & $\left[ 0.442, 0.458\right]$ \\ \hline
$\delta$ & $0$ & $\left[-0.2, 0.2\right]$ \\ \hline
& $\pi/2$  & $\left[1.37, 1.77\right]$ \\ \hline
$\sin^2\left(2\theta_{13}\right)$ & $0.085$ & $\left[0.080, 0.090\right]$ \\ \hline
$\Delta m_{31}^2$ & $2.457 \times 10^{-3}$ eV$^2$ & $\left[ 2.447, 2.467\right] \times 10^{-3}$ eV$^2$ \\ \hline
\end{tabular}
\caption{Expected measurement precision at DUNE for parameters of interest assuming physical values listed. We note here that the measurement precision of DUNE for $\delta$ is mostly independent of its physical value, however we list the precision assuming $\delta = 0$ or $\pi/2$ here, as projected by the DUNE collaboration.}
\label{tab:Params}
\end{center}
\end{table}
This assumes a total exposure of $300$ kt-MW-years, consistent with experimental expectations.

\begin{figure}[!htbp]
\centering
\includegraphics[width=\linewidth]{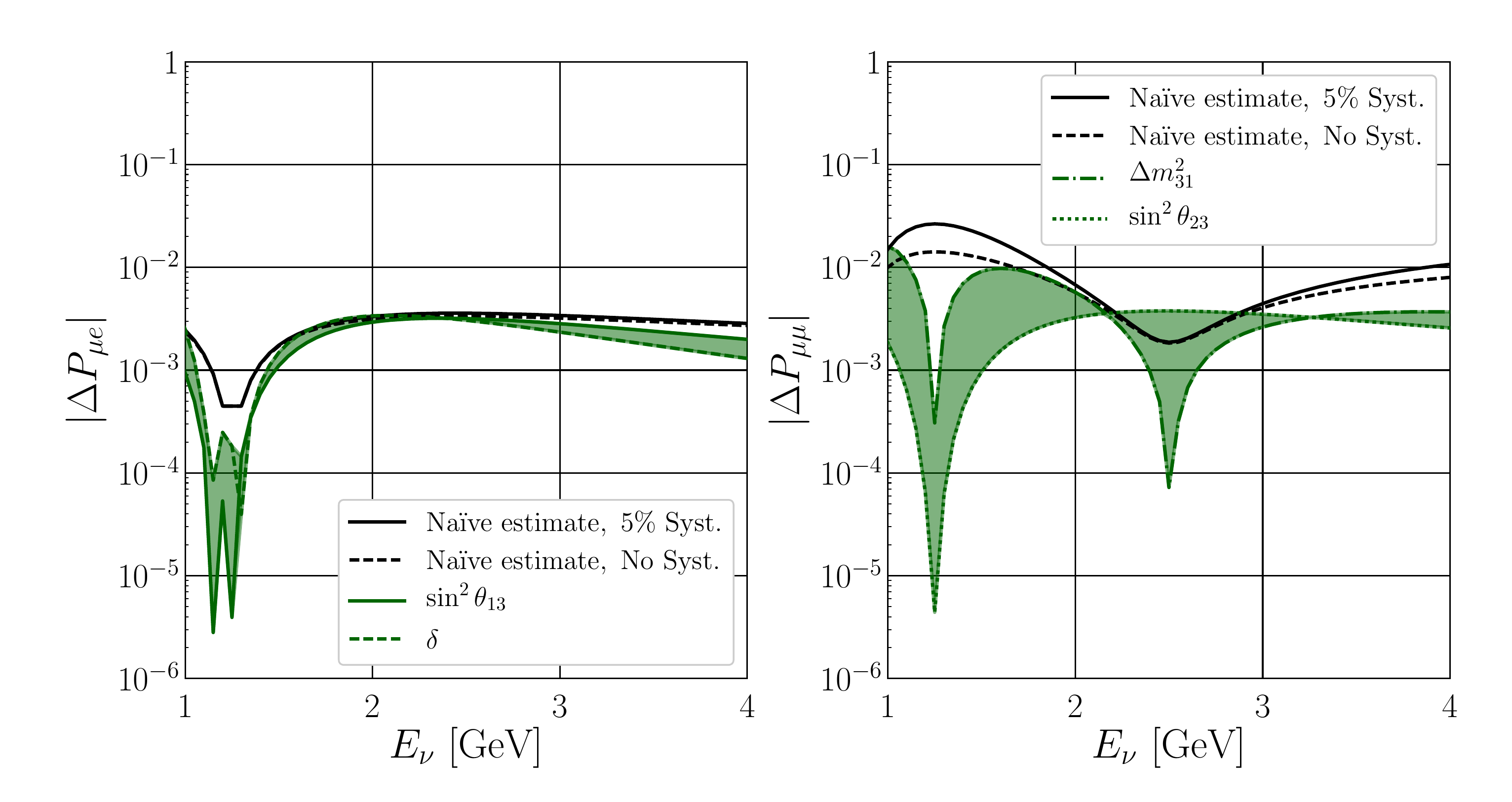}
\caption{\textbf{Black} lines: na{\"i}ve prediction of measurement precision of oscillation probability $|\Delta P_{\alpha\beta}|$ assuming $N_1 = 10^3$ unoscillated events for all energies. Solid \textbf{black} lines include a 5\% uncorrelated bin-to-bin systematic uncertainty, where dashed \textbf{black} lines are for only statistical uncertainties. We have included a bin-to-bin measurement improvement factor of $\sqrt{5}$ to this na{\"i}ve estimate as discussed in the text. \textbf{{\color{paramcol} Green}}: Change to oscillation probabilities while changing oscillation parameters between their central values and $\pm 1\sigma$ extremes, as given in Table~\ref{tab:Params}. In the left panel, we show the impact on appearance probability $P_{\mu e}$ for the parameters measured (predominantly) by this channel, $\sin^2\theta_{13}$ (solid) and $\delta$ (dashed). In the right panel, we show the impact on disappearance probability $P_{\mu\mu}$ and its associated parameters, $\Delta m_{31}^2$ (dot-dashed) and $\sin^2\theta_{23}$ (dotted). We do not display antineutrino probability precisions here, but the result is qualitatively the same.}
\label{fig:ParamsAndNaive}
\end{figure}

The appearance channel $P_{\mu e}$ (and its CP-conjugate) has sensitivity predominantly to the parameters $\sin^2\theta_{13}$ and $\delta$, where the disappearance channel $P_{\mu\mu}$ has sensitivity to $\sin^2\theta_{23}$ and $\Delta m_{31}^2$. With this in mind, we calculate the oscillation probability for a given channel assuming the physical values listed in Table~\ref{tab:Params}, as well as the oscillation probability with the parameter at its $\pm 1\sigma$ value. We calculate the change in probability between these two and show this in Fig.~\ref{fig:ParamsAndNaive}. 
We show only the impact of $\sin^2\theta_{13}$ and $\delta$ in the appearance\footnote{The experimental sensitivity to the CP-violating phase $\delta$ comes largely from comparing neutrino and antineutrino appearance channels. Here, we simply display the change to the neutrino oscillation probability from changing $\delta$, but insist that this is an incomplete picture of the experimental sensitivity.} panel (left) and $\sin^2\theta_{23}$ and $\Delta m_{31}^2$ in the disappearance panel (right). Additionally, we include the na{\"i}ve estimates with and without 5\% systematic uncertainties discussed in Section~\ref{subsec:Naive}. We see here that the na{\"i}ve estimate with a $\sqrt{5}$ bin-to-bin measurement improvement factor comes close\footnote{The fact that the changes in oscillation probability induced by changing parameters is, for some energies, significantly lower than our na{\"i}ve estimate implies that the parameters are being measured where $|\Delta P_{\alpha\beta}|$ is largest, e.g. near $1.6$ GeV for $\Delta m_{31}^2$ in the disappearance channel (Fig.~\ref{fig:ParamsAndNaive}, left panel, dot-dashed line).} to capturing the true sensitivity to oscillation probability changes that comes from changing oscillation parameters. The necessary change to the oscillation probability in order to be measured at DUNE is on the level of $2\times 10^{-3}$ (greater than $2 \times 10^{-3}$) for the appearance (disapperance) channel.

\subsection{Matter Density Profile Effects}
\label{subsec:MatterDensity}
Recently, Ref.~\cite{Roe:2017zdw} studied different models of the Earth's matter density profile and the resulting density as a function of distance between Fermilab and the future location of the DUNE detector, in South Dakota. The models discussed in detail are Shen-Ritzwoller~\cite{Shen:2016xxx}, Crustal~\cite{2013EGUGA..15.2658L}, and PEMC~\cite{Pemc:1975xxx}. The author of Ref.~\cite{Roe:2017zdw} cautions that these different matter density models lead to changes in oscillation probabilities for the energy range of interest at DUNE.
\begin{figure}[!htbp]
\centering
\includegraphics[width=0.5\linewidth]{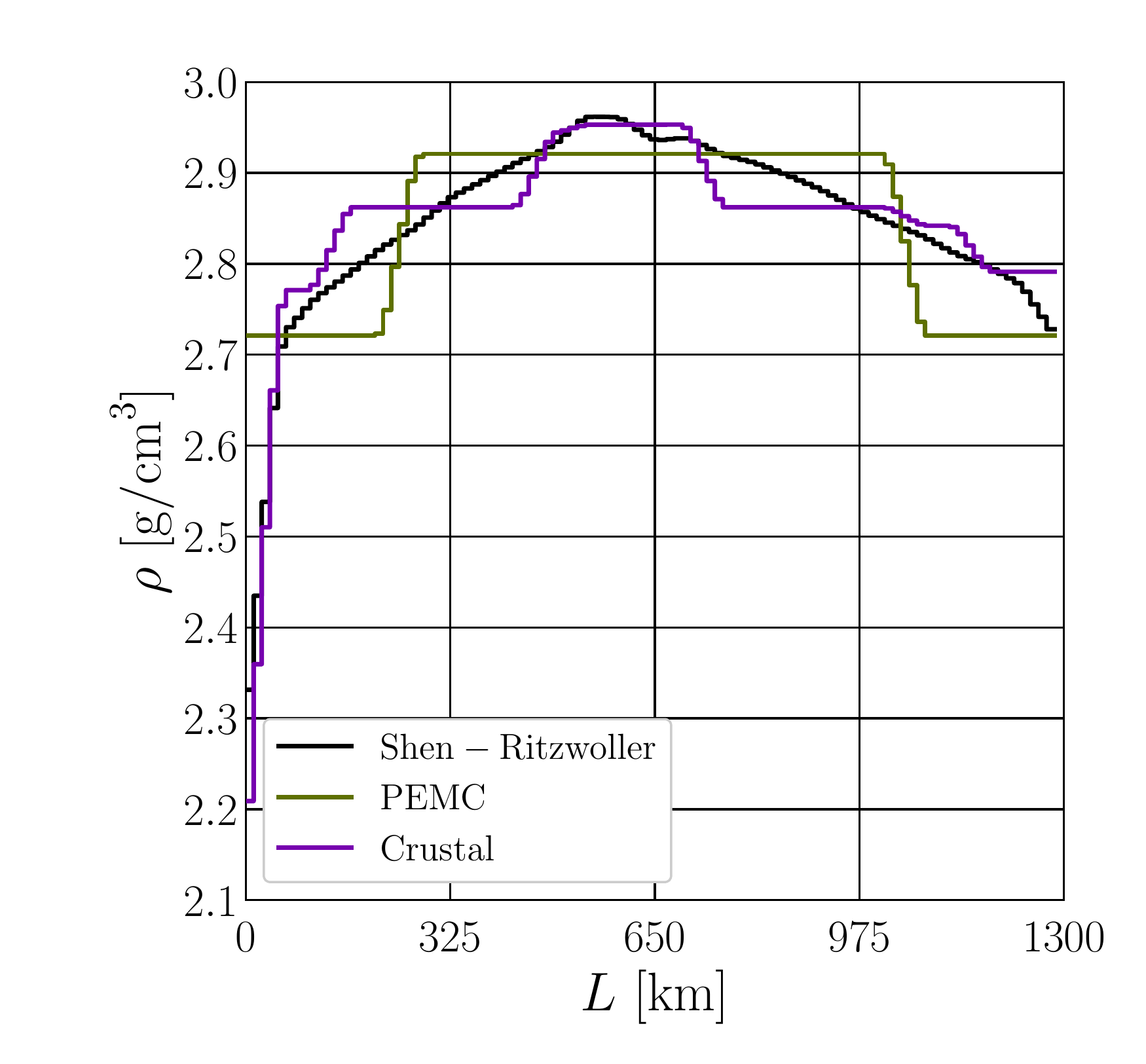}
\caption{The density maps considered here, given in Ref.~\cite{Roe:2017zdw} and scaled such that $\rho_{\mathrm{Avg.}} = 2.845$ g/cm$^3$. Each density map is divided into $N = 100$ segments.}
\label{fig:DensityMaps}
\end{figure}
In Fig.~\ref{fig:DensityMaps}, we reproduce the Shen-Ritzwoller, PEMC, and Crustal maps considered in Ref.~\cite{Roe:2017zdw}, all normalized to the same average density $\rho_\mathrm{Avg.} = 2.845$ g/cm$^3$. The profiles have been divided into $N = 100$ piecewise constant segments.

\begin{figure}[!htbp]
\centering
\includegraphics[width=\linewidth]{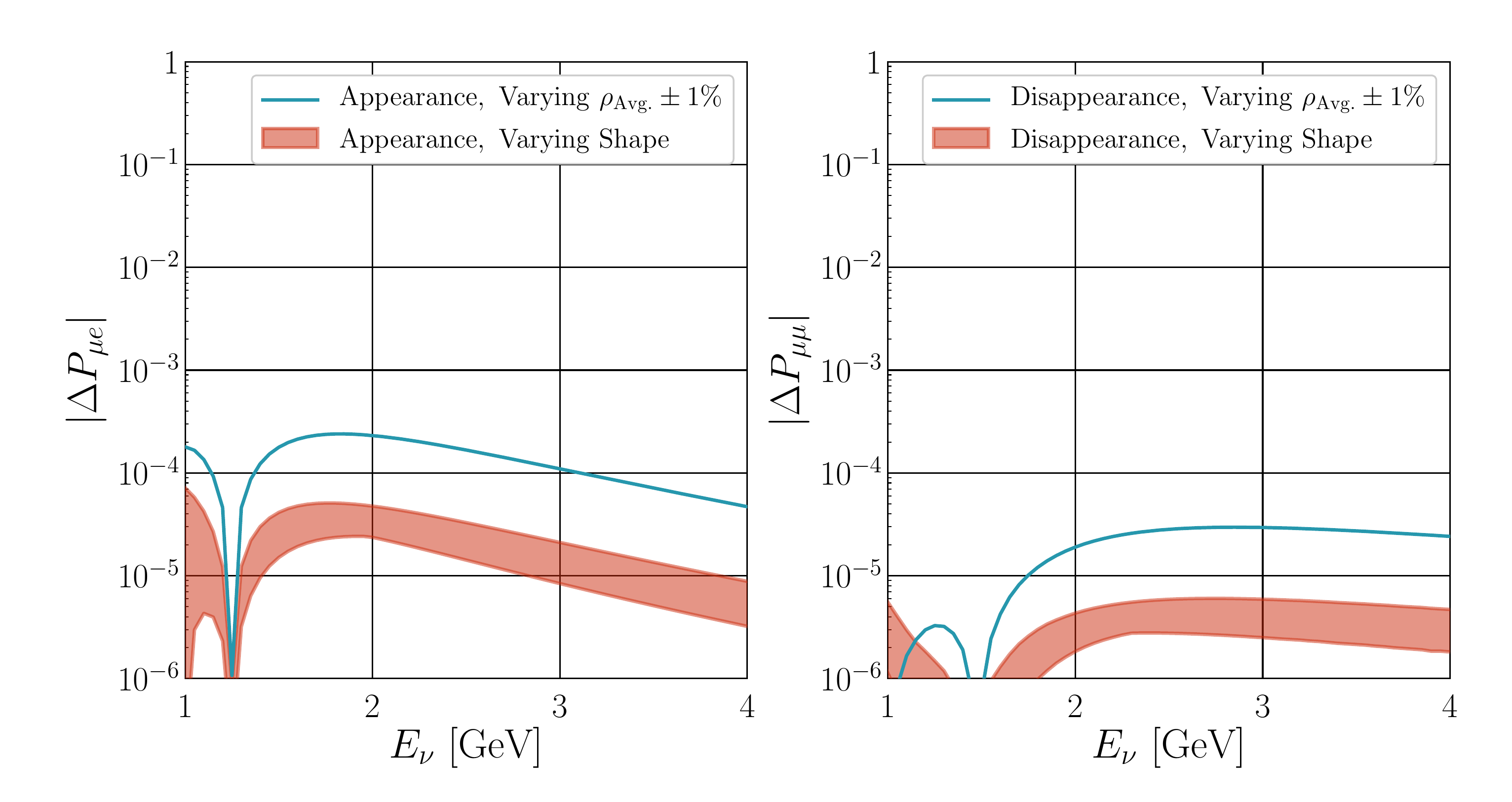}
\caption{Shaded regions (\textbf{{\color{shapecol} red}}): the range of change in oscillation probabilities obtained while changing the shape of the matter density profile, comparing the Crustal map, Shen-Ritzwoller map, and the PEMC map, as detailed in Ref.~\cite{Roe:2017zdw}. Matter density profiles have been normalized so that $\rho_{\mathrm{Avg.}} = 2.845$ g/cm$^3$. Solid lines (\textbf{{\color{avgcol} blue}}) display the change in oscillation probabilities when changing a constant matter density by $\pm 1\%$ of $\rho_\mathrm{Avg.} = 2.845$ g/cm$^3$. The left panel displays change in appearance probability $P_{\mu e}$ where the right panel displays change in disappearance probability $P_{\mu\mu}$, both for neutrino oscillation.}
\label{fig:ShapeNorm}
\end{figure}

In this subsection, we consider changes to the oscillation probability due to these different matter density profiles. We separate this discussion into probability differences induced by changes in the density profile shape and those induced by changes in the average density. First, we calculate the oscillation probabilities with identical oscillation parameters for all three density profiles (with $N=100$ regions) as well as $\rho_\mathrm{Avg.} = 2.845$ g/cm$^3$. We then calculate the differences between probabilities for each pair of density profiles, and show the range of differences obtained by this process in Fig.~\ref{fig:ShapeNorm} in the \textbf{{\color{shapecol} red}} shaded regions.

Next, we calculate, for a flat matter density profile, the change in oscillation probabilities when $\rho_\mathrm{Avg.}$ is changed between $2.845$ g/cm$^3$ and $\pm 1\%$, or $[2.82, 2.87]$ g/cm$^3$. The difference between the upper and lower range of this is negligible, and we display the resulting difference in probability $|\Delta P_{\alpha\beta}|$ also in Fig.~\ref{fig:ShapeNorm} as a solid \textbf{{\color{avgcol} blue}} line. We see that a $1\%$ change in the average matter density induces changes to the probability nearly an order of magnitude larger than changes in shape that are $\mathcal{O}(10\%)$ of the average density locally. Moreover, we see that both of these effects generate changes in the probability that are \textit{far} below what is necessary at DUNE to be measurable. Additionally, we see that the impact on the disappearance channel $P_{\mu\mu}$ is lower than that for the appearance channel $P_{\mu e}$ for all energies of interest. This is due to the fact that matter effects impact the appearance channel more significantly than the disappearance channel when comparing with vacuum oscillation probabilities.

While the density profiles here are of particular interest for DUNE, we additionally would like to know whether this behavior -- that the impact of changing the average density dwarfs changing the shape of the profile -- is generic. In Appendix~\ref{sec:AppendixShapeNorm}, we consider a simple matter density profile that has two free parameters, one that governs the shape of the distribution, and one that governs its average density. We show that in general, a fractional change to the shape leads to probability differences that are five times smaller than those induced by the same fractional change in the average density.

Here, we have considered changes to the average density $\rho_\mathrm{Avg.}$ at the level of $1\%$, in agreement with the largest uncertainties on $\rho_\mathrm{Avg.}$ discussed in Ref.~\cite{Roe:2017zdw}. Clearly, uncertainties at this level will have no impact at measurable levels at DUNE. In order to see how well DUNE can measure $\rho_\mathrm{Avg.}$ without any prior information, we allow it to be a free parameter in a fit in Appendix~\ref{sec:MeasureRho}. There, we see that DUNE requires changes to the average density on the order of $25\%$ to make a measurable impact. We also see that allowing a $1\%$ prior on $\rho_\mathrm{Avg.}$ has no impact on the measurement of any oscillation parameters. Even without a prior, the only parameter measurement that worsens is $\delta$, however it is a small effect.

\subsection{Perturbative Approaches}
\label{subsec:PerturbativeSensitivity}
Constructing oscillation probabilities for three-neutrino oscillations in the presence of matter has been of interest for several decades~\cite{Zaglauer:1988gz,Sato:1997st,Arafune:1997hd,Minakata:1998bf,Cervera:2000kp,Freund:2001pn,Kimura:2002wd,Akhmedov:2004ny,Blennow:2013rca,Minakata:2015gra,Denton:2016wmg,Li:2016pzm}. Here, we focus on a method~\cite{Denton:2016wmg,Denton:2018hal} specifically developed for calculating oscillation probabilities perturbatively for long-baseline experiments such as DUNE. This approach, which we will refer to as the DMP method, provides a much faster way of calculating a probability, compared with that discussed above, which relies on calculating $N$ $3\times 3$ matrix exponentials for each neutrino energy considered. In the DMP method, one calculates changes to the mixing angles $\theta_{13} \to \widetilde{\theta}_{13}$ and $\theta_{12} \to \widetilde{\theta}_{12}$, as well as changes to the mass-splittings $\Delta m_{ji}^2 \to \Delta \widetilde{m}_{ji}^2$. We will be concerned with the zeroth order expansion of the DMP method, and reproduce the results here for completeness.

The modifications depend on a combination of the (unperturbed) mass-splittings, $\Delta m_{ee}^2 \equiv \cos^2\theta_{12} \Delta m_{31}^2 + \sin^2\theta_{12} \Delta m_{32}^2$. The modifications to the mixing angles then, are
\begin{align}
\cos{2\widetilde{\theta}_{13}} &= \frac{\left(\cos{2\theta_{13}} - a/\Delta m_{ee}^2\right)}{\sqrt{\left(\cos{2\theta_{13}} - a/\Delta m_{ee}^2\right)^2 + \sin^2 2\theta_{13}}}, \\
\cos{2\widetilde{\theta}_{12}} &= \frac{\left(\cos{2\theta_{12}} - a'/\Delta m_{21}^2\right)}{\sqrt{\left(\cos{2\theta_{12}} - a'/\Delta m_{21}^2\right)^2 + \sin^2 2\theta_{12} \cos^2{\left(\widetilde{\theta}_{13} - \theta_{13}\right)}}},
\end{align}
where $a' \equiv a\cos^2\widetilde{\theta}_{13} + \Delta m_{ee}^2 \sin^2{\left(\widetilde{\theta}_{13} - \theta_{13}\right)}$. Both $\widetilde{\theta}_{12}$ and $\widetilde{\theta}_{13}$ are in the range $[0, \pi/2]$.

The modified mass-splittings are
\begin{align}
\Delta \widetilde{m}_{21}^2 &= \Delta m_{21}^2 \sqrt{\left(\cos{2\theta_{12}} - a'/\Delta m_{21}^2\right)^2 + \sin^2 2\theta_{12} \cos^2{\left(\widetilde{\theta}_{13} - \theta_{13}\right)}}, \\
\Delta \widetilde{m}_{31}^2 &= \Delta m_{31}^2 + \frac{1}{2}\left(2a - 3a' + \Delta \widetilde{m}_{21}^2 - \Delta m_{21}^2\right).
\end{align}
With these perturbative angles and mass-splittings, the zeroth-order probability can be calculated in the DMP scheme\footnote{In Refs.~\cite{Denton:2016wmg,Denton:2018hal}, the mixing matrix used is distinct from the PDG convention in Ref.~\cite{Patrignani:2016xqp}. These differences are not realizable at zeroth order of perturbation.}. The method for calculating higher-order perturbative corrections can be found in Refs.~\cite{Denton:2016wmg,Denton:2018hal}.

We can compare the zeroth-order DMP probabilities with those calculated using a proper matrix exponential for a constant matter density and $N=1$ layer in order to characterize how precise the DMP method is.
\begin{figure}
\centering
\includegraphics[width=\linewidth]{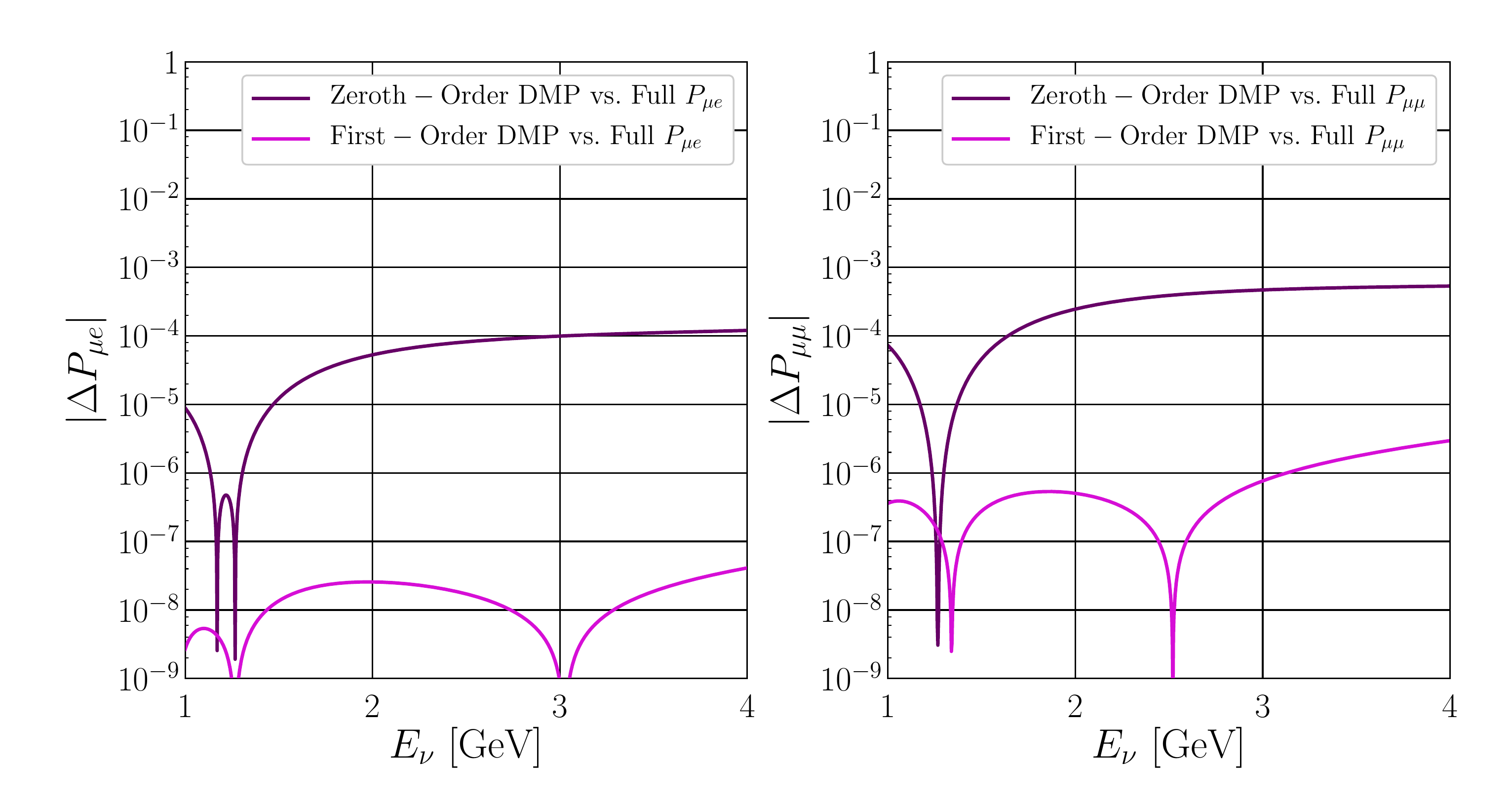}
\caption{Change in oscillation probabilities between the DMP perturbative method at zeroth-order (\textbf{{\color{pertcol} purple}}) and first-order (\textbf{{\color{firstcol} pink}}) discussed in Section~\ref{subsec:PerturbativeSensitivity} and a matrix-exponential-calculated oscillation probability assuming one layer, both with constant matter density $\rho = 2.845$ g/cm$^3$. The left panel displays differences for appearance channel neutrino oscillation probabilities $P_{\mu e}$, and the right panel displays differences for disappearance channel probabilities $P_{\mu\mu}$.}
\label{fig:DMP}
\end{figure}
The differences in oscillation probability are shown in Fig.~\ref{fig:DMP} in \textbf{{\color{pertcol} purple}}. Additionally, we include the first-order DMP probabilities in \textbf{{\color{firstcol} pink}}. We see that the resulting $|\Delta P_{\alpha\beta}|$, even at zeroth-order is well below the range necessary for detection at DUNE, implying that the zeroth-order DMP approach is sufficient for calculating oscillation probabilities for DUNE. The vertical axis in Fig.~\ref{fig:DMP} extends far lower than those in Figs.~\ref{fig:ParamsAndNaive} and \ref{fig:ShapeNorm} in order to display the first-order precision. In order for DUNE to be sensitive to this level of $|\Delta P_{\alpha\beta}|$, at least two orders of magnitude larger statistics would be necessary, as discussed in Section~\ref{subsec:Naive}.

We encourage the use of this approach, as compiled zeroth-order DMP {\sc C++} code can calculate an oscillation probability in $\mathcal{O}(10^{-7})$ s, whereas compiled {\sc C++} code with the full matrix exponential (even for $N=1$ layer of matter) calculates a probability in $\mathcal{O}(10^{-5})$ s. A factor of $100$ faster calculation can drastically reduce computation time for large parameter spaces. If one requires the precision demonstrated by the first-order DMP method shown in Fig.~\ref{fig:DMP}, the amount of time to calculate a probability is not significantly longer than at zeroth-order.


\setcounter{footnote}{0}
\setcounter{equation}{0}
\section{Discussion and Conclusions}
\label{sec:Conclusions}
\begin{figure}[!htbp]
\centering
\includegraphics[width=\linewidth]{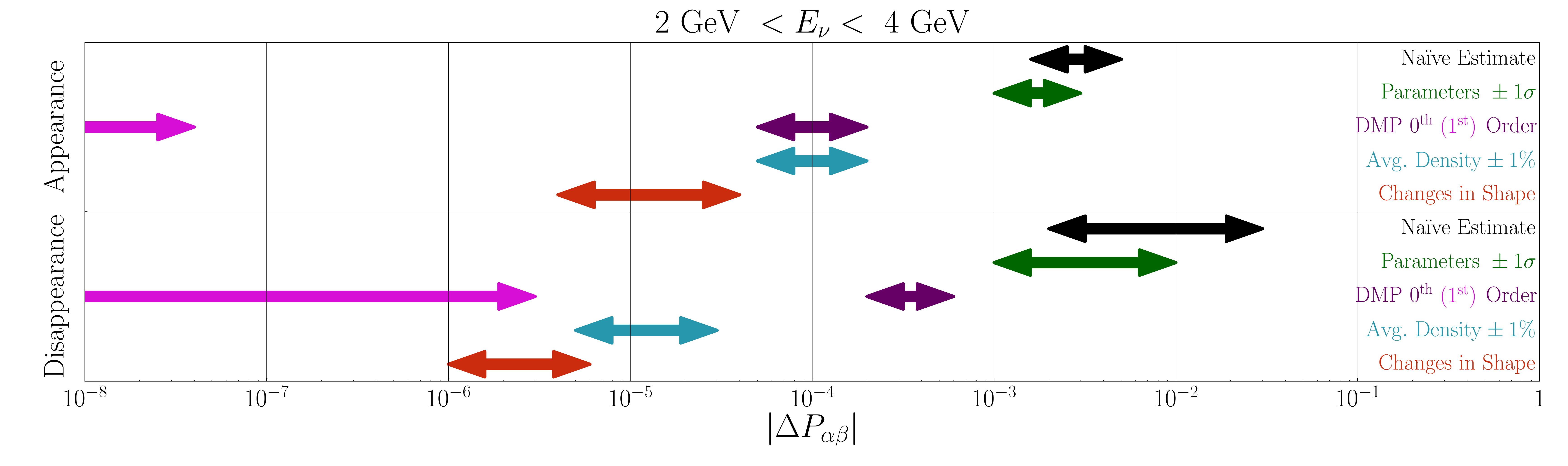}
\caption{A summary of the scales of $|\Delta P_{\mu e}|$ (appearance, top) and $|\Delta P_{\mu\mu}|$ (disappearance, bottom) discussed in this paper: na{\"ive} sensitivity estimates in Section~\ref{subsec:Naive} (\textbf{black}), differences from parameter changes in Section~\ref{subsec:OscParams} (\textbf{{\color{paramcol} green}}), changes to the matter density profile average density (\textbf{{\color{avgcol} blue}}) and shape (\textbf{{\color{shapecol} red}}) discussed in Section~\ref{subsec:MatterDensity}, and the precision of the zeroth-order (\textbf{{\color{pertcol} purple}}) and first-order (\textbf{{\color{firstcol} pink}}) DMP perturbative approach from Section~\ref{subsec:PerturbativeSensitivity}. We restrict the neutrino energy to be in the range $2$ GeV $< E_\nu <$ $4$ GeV, where the number of unoscillated events is highest for each channel.}
\label{fig:Scales}
\end{figure}

In this manuscript, we have analyzed the impact of matter effects at the Deep Underground Neutrino Experiment, and shown that the only significant quantity regarding these, for the sake of measuring neutrino oscillation parameters, is the average density $\rho_\mathrm{Avg.}$. We have estimated the sensitivity to differences in oscillation probabilities at DUNE for both appearance and disappearance channels using both a na{\"i}ve statistical approach and analyzing the experiment's sensitivity to oscillation parameters. Additionally, we have shown that differences to the oscillation probability caused by changing the average density within its allowed region (or even inflated significantly) are smaller than those required for DUNE to detect~\cite{Roe:2017zdw}. The perturbative approach in Refs.~\cite{Denton:2016wmg,Denton:2018hal} can calculate oscillation probabilities precisely enough to capture all measurable effects at DUNE. Changes in the matter density profile shape, e.g. the three profiles considered in Ref.~\cite{Roe:2017zdw}, induce changes to the oscillation probability smaller than all of these and will be immeasurable at DUNE. A summary of these scales of $|\Delta P_{\alpha\beta}|$ for both appearance and disappearance channels is shown in Fig.~\ref{fig:Scales}.

Additionally, we have also explored the computation time saved by using the perturbative approach as opposed to a more exact calculation time; several orders-of-magnitude faster. Because of this, we encourage the use of such perturbative approaches, as the precision is more than capable of calculating all detectable oscillation probabilities at DUNE.

Briefly, we discuss these results in a broader context. Many studies of upcoming long-baseline neutrino experiments consider the possibility that neutrinos have additional interactions with the matter along the path of propagation, dubbed non-standard neutrino interactions (NSI) (see, e.g., Refs.~\cite{Masud:2015xva,deGouvea:2015ndi,Coloma:2015kiu,Liao:2016hsa,Masud:2016bvp,Masud:2016gcl,Blennow:2016etl,Deepthi:2016erc,Liao:2016orc,Ghosh:2017lim,Farzan:2017xzy,Deepthi:2017gxg} for discussions of NSI at DUNE). These scenarios are testable at DUNE, in large part due to the matter effects discussed in this manuscript. Because these NSI alter the interaction potential discussed in Section~\ref{sec:Oscillations}, the DMP perturbative expansion cannot be used here. Other perturbative methods exist for these scenarios, such as that detailed in Ref.~\cite{Kikuchi:2008vq}, however they do not offer the same level of precision as the DMP method. Regardless of the calculation method considered for NSI, we still note that the effects due to changing matter density profile shape are subdominant to any measurable impacts at DUNE. In addition to parameter degeneracies discussed in Appendix~\ref{sec:MeasureRho}, further degeneracies will exist in studying NSI if one considers a changing average matter density (particularly $\rho_\mathrm{Avg.}$ vs. $\epsilon_{ee}$).

In Appendix~\ref{sec:SystAppendix}, we explored the modifications to our na{\"i}ve sensitivity estimate in light of systematic uncertainties. We saw that, as expected, systematic uncertainties only make measurements more difficult, and that at least two order of magnitude more events are necessary to improve a measurement by an order of magnitude.

We analyzed a simple matter density profile model in Appendix~\ref{sec:AppendixShapeNorm} in order to explore whether, in general, changes to a matter density profile shape matter far less than changes to the average density. We found this to be the case, in agreement with the discussion in Section~\ref{subsec:MatterDensity}. Additionally, we found that changes to the average density on the order of $\pm 25\%$ its true value are required to make measurable changes to the oscillation probability at DUNE.

Finally, we explored the capability of DUNE to measure $\rho_\mathrm{Avg.}$ assuming a constant matter density profile in Appendix~\ref{sec:MeasureRho}. The results here agreed with those in Appendix~\ref{sec:AppendixShapeNorm}, DUNE will be able to independently measure $2.5$ g/cm$^3$ $\lesssim \rho_\mathrm{Avg.} \lesssim 3.5$ g/cm$^3$ at a $1\sigma$ level, and this statement is true regardless of the true value of $\delta$. We also saw that the only oscillation parameter with its measurement impacted by a free parameter $\rho_\mathrm{Avg.}$ is $\delta$, and even so it is not a large effect.

\section*{Acknowledgements}
The work of KJK is supported in part by DOE grant \#de-sc0010143. We acknowledge the use of the Quest computing cluster at Northwestern University for a portion of this research. KJK thanks the Neutrino Physics Center at Fermilab for providing support during the completion of this work.

This manuscript has been authored by Fermi Research Alliance, LLC under Contract No. DE-AC02-07CH11359 with the U.S. Department of Energy, Office of Science, Office of High Energy Physics.

This project has received funding/support from the European Union's Horizon 2020 research and innovation programme under the Marie Sklodowska-Curie grant agreement No 690575.
 This project has received funding/support from the European Union's Horizon 2020 research and innovation programme under the Marie Sklodowska-Curie grant agreement No 674896.

\appendix
\section{Effect of Systematic Uncertainty on Measurement Precision}\label{sec:SystAppendix}
In Section~\ref{sec:Precision} and Fig.~\ref{fig:PrecisionStat}, we discussed the fact that in order to be an order of magnitude more precise in measuring $|\Delta P_{\alpha\beta}|$ or $|\Delta P_{\alpha\beta}|/P_{\alpha\beta}$, an experiment required two orders of magnitude larger statistics. Here, we repeat the exercise assuming an uncertainty associated with $N_1$, the product of fluxes, cross-sections, and efficiencies. 
\begin{figure}[!htbp]
\centering
\includegraphics[width=0.6\linewidth]{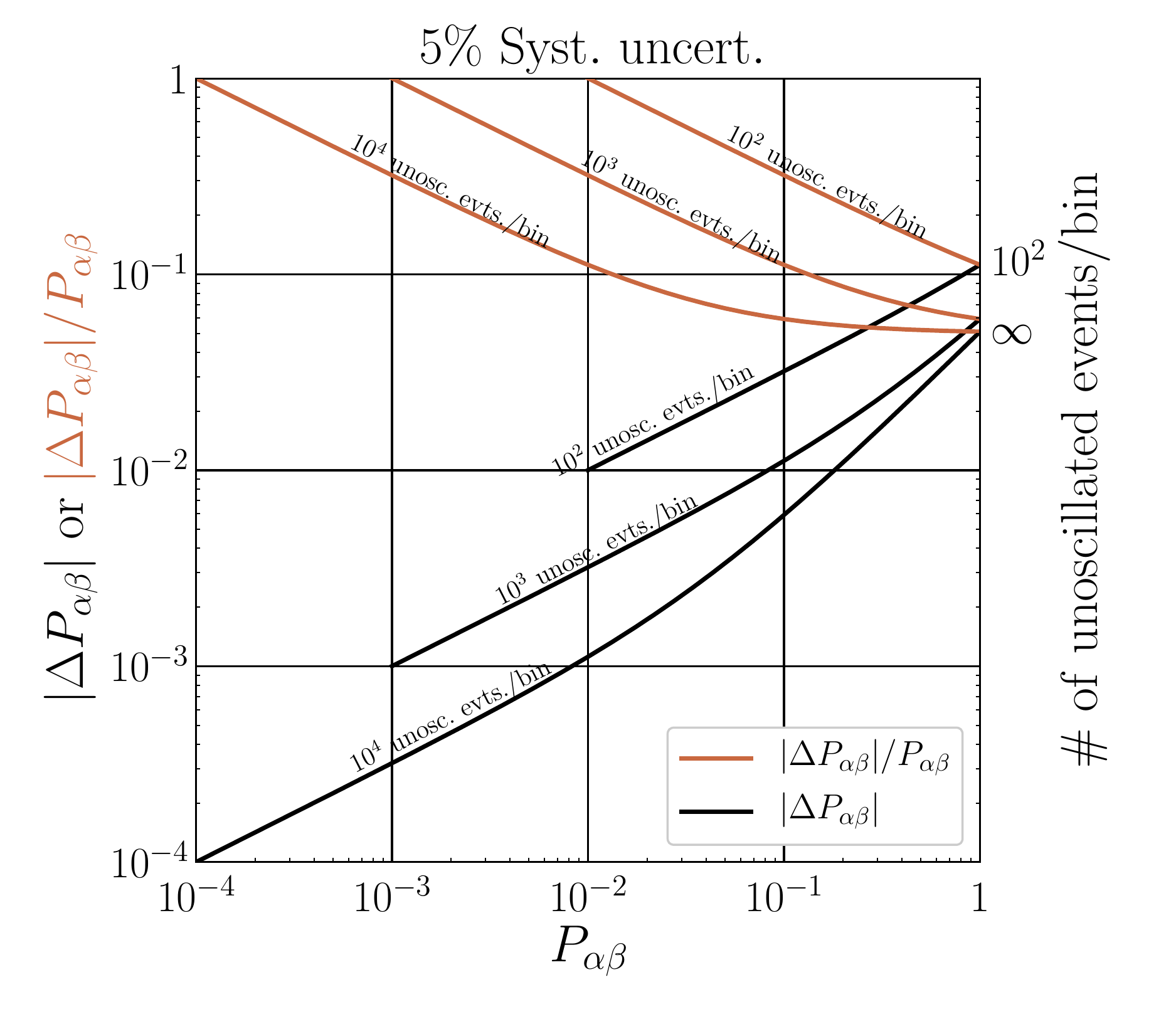}
\caption{Measurement precision of a single-bin experiment with statistical uncertainty and $5\%$ normalization uncertainty in the bin as a function of the oscillation probability $P_{\alpha\beta}$. \textbf{{\color{dpcol} Orange}} lines give precision in terms of fractional uncertainty $|\Delta P|/P$ where solid lines give precision in terms of $|\Delta P|$. The right axis, along with annotations, denotes the number of unoscillated events necessary in a bin to attain the given precision.}
\label{fig:PrecisionsSyst}
\end{figure}
With the number of events $N = N_1 \times P_{\alpha\beta}$, we can solve for $P_{\alpha\beta}$ and calculate the fractional uncertainty on $P_{\alpha\beta}$, $|\Delta P_{\alpha\beta}|/P_{\alpha\beta}$ as
\begin{equation}
\frac{|\Delta P_{\alpha\beta}|}{P_{\alpha\beta}} = \sqrt{\frac{1}{N_1 \times P_{\alpha\beta}} + \left(\frac{\sigma_{N_1}}{N_1}\right)^2},\label{eq:SystPrecision}
\end{equation}
where $\sigma_{N_1}$ is the uncertainty on $N_1$. We assume that $\sigma_{N_1}/N_1 \simeq 5\%$. In this limit, the statistical uncertainty on $N_1$ sets a lower limit on $|\Delta P_{\alpha\beta}|/P_{\alpha\beta}$ of $\sigma_{N_1}/N_1$, evident in Fig.~\ref{fig:PrecisionsSyst} in that the curves tend towards a single point at $P_{\alpha\beta} = 1$, even as the number of unoscillated events increases. The same conclusions drawn in discussing Fig.~\ref{fig:PrecisionStat} hold here: a factor of 100 increase in statistics improves sensitivity to $|\Delta P_{\alpha\beta}|$ or $|\Delta P_{\alpha\beta}|/P_{\alpha\beta}$ by at most a factor of 10.

\section{Two-layer Model: Estimate of Shape/Normalization Sensitivity}\label{sec:AppendixShapeNorm}
\begin{figure}[!htbp]
\centering
\includegraphics[width=0.4\linewidth]{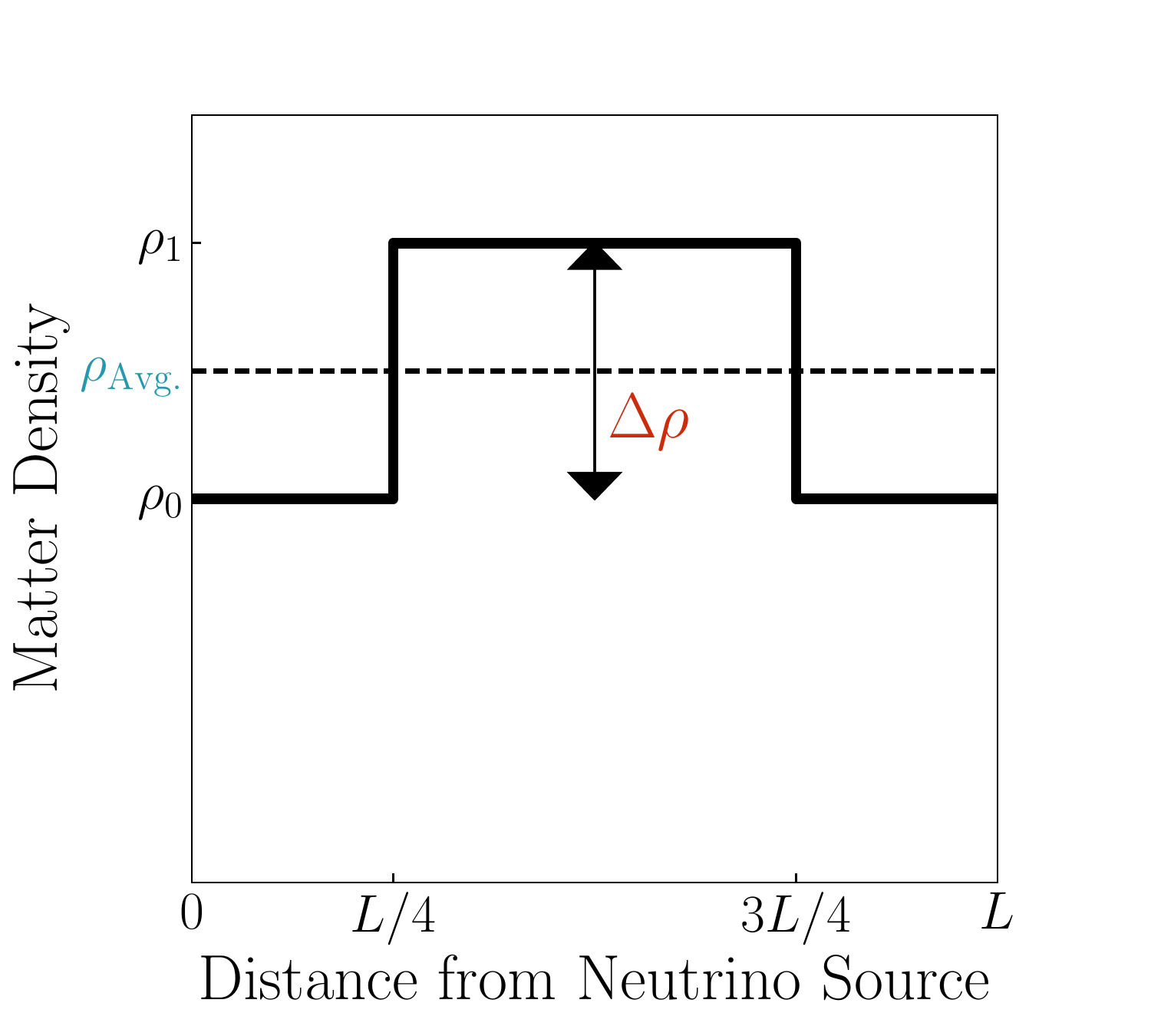}
\caption{Simple, two-layer matter density profile with two free parameters, either $\rho_0$ and $\rho_1$ or $\rho_\mathrm{Avg.} = (\rho_0 + \rho_1)/2$ and $\Delta \rho = \rho_1 - \rho_0$. For simplicity, this model assumes the density profile is symmetric, and that the middle, raised portion has the same distance as the two outside portions combined.}
\label{fig:TwoLayer}
\end{figure}

In general, the matter density profiles considered in this work and Ref.~\cite{Roe:2017zdw} are roughly symmetric over the baseline of DUNE. With this as motivation, we construct a simplified two-layer density model as shown in Fig.~\ref{fig:TwoLayer}.
This model has two free parameters, $\rho_\mathrm{Avg.}$ and $\Delta \rho$, the average matter density and the size of the middle step, respectively\footnote{The individual matter densities can be written as $\rho_0 \equiv \rho_\mathrm{Avg.} - \Delta \rho/2$ and $\rho_1 \equiv \rho_\mathrm{Avg.} + \Delta \rho/2$.}. We use these as our free parameters to separate effects due to shape from those due to normalization of the density profile. 

\begin{figure}[!htbp]
\centering
\includegraphics[width=\linewidth]{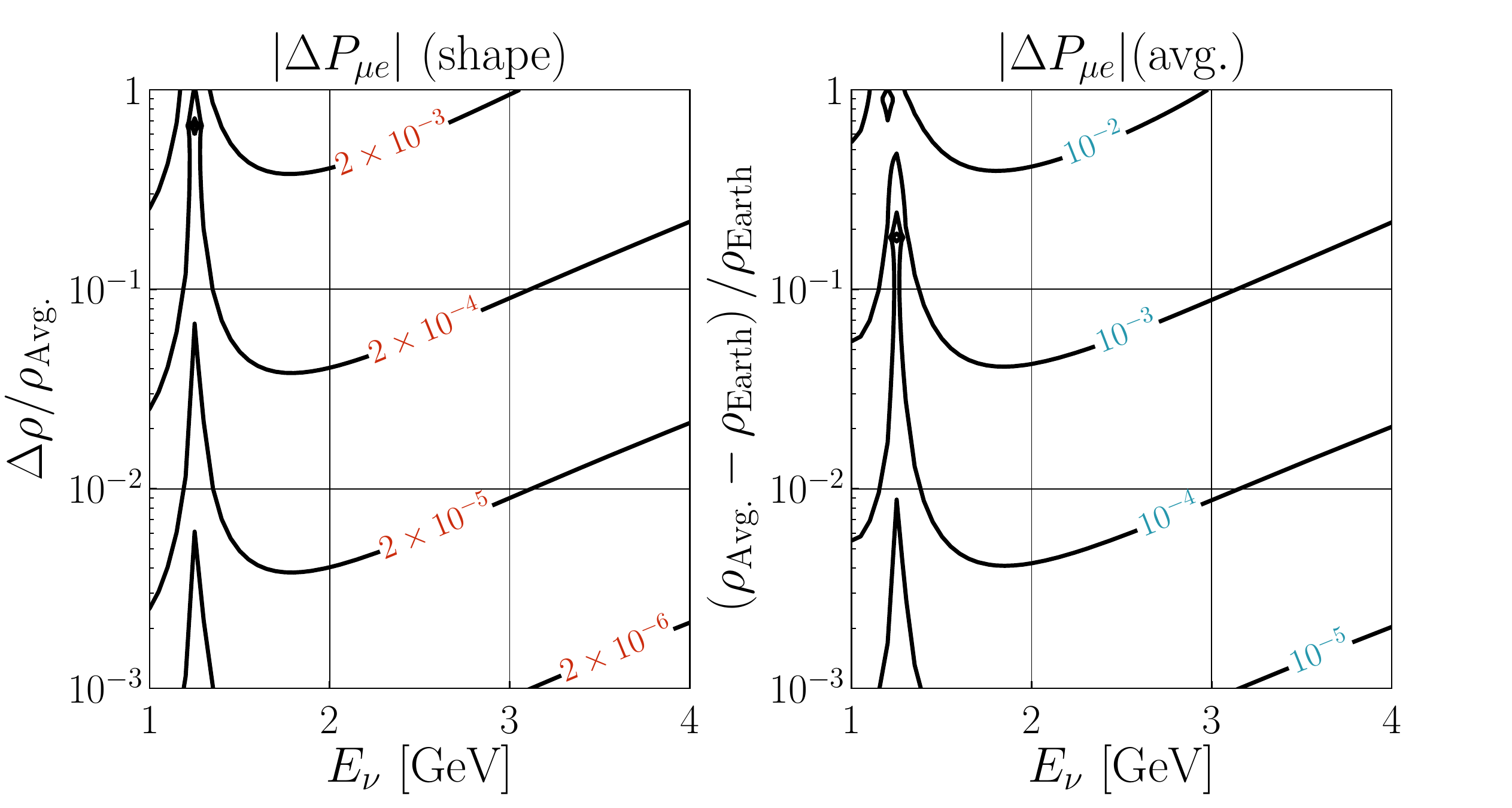}
\caption{Change in oscillation probability $P_{\mu e}$ when changing the shape (left) or average density (right) of the matter density profile discussed in Appendix~\ref{sec:AppendixShapeNorm} and shown in Fig.~\ref{fig:TwoLayer}. In the left panel, we keep $\rho_\mathrm{Avg.} = 2.845$ g/cm$^3$ fixed and vary $\Delta \rho$, where in the right panel, we keep $\Delta \rho = 0$ fixed, and vary $\rho_\mathrm{Avg.}$. Black lines indicate contours of constant $|\Delta P_{\mu e}|$, with labels indicating contours of interest.}
\label{fig:TwoLayerShapeNorm}
\end{figure}

Using $\rho_\mathrm{Earth} \equiv 2.845$ g/cm$^3$ as our benchmark for comparison, we analyze the effect of independently changing $\Delta \rho$ (left) and $\rho_\mathrm{Avg.}$ (right) on the oscillation probability $P_{\mu e}$ in Fig.~\ref{fig:TwoLayerShapeNorm}.
We allow for different neutrino energies $E_\nu$ as well. While this is a na\"ive approximation to the matter density profile of Earth, we can draw several conclusions. First, for a similar percentage change in the shape difference as for the average difference, e.g. $\Delta \rho/\rho_\mathrm{Avg.} = \left(\rho_\mathrm{Avg.} - \rho_\mathrm{Earth}\right)/\rho_\mathrm{Earth} = 10^{-1}$, the effect of changing the shape is roughly a factor of five smaller than that of changing the average density. This reinforces our claim in Section~\ref{sec:Precision} that (large) changes to the average density can be measurable at DUNE where the changes in shape can not. Second, we see that changes on the order of $\Delta \rho/\rho_\mathrm{Avg.} \simeq 10^{-1} - 1$ are necessary for shape changes to cause probability differences on the order of $10^{-3} - 10^{-2}$, the level discussed as necessary for being measured at DUNE. No profile discussed in Ref.~\cite{Roe:2017zdw} has changes on this level. Additionally, for changes to the average density of $\left(\rho_\mathrm{Avg.} - \rho_\mathrm{Earth}\right)/\rho_\mathrm{Earth}$ on the level of a few times $10^{-1}$, we achieve the change in probability necessary for sensitivity, which agrees with the more thorough analysis of Appendix~\ref{sec:MeasureRho}.

\section{Measurement of Matter Density}\label{sec:MeasureRho}
\setcounter{equation}{0}

We use the DMP~\cite{Denton:2018hal} method for calculating oscillation probabilities at zeroth order, as well as code developed to simulate expected event yields at DUNE (see Refs.~\citep{Berryman:2015nua,deGouvea:2015ndi,deGouvea:2016pom} for further explanation of this code), to estimate the ability of DUNE to measure oscillation parameters. We use the Markov Chain Monte Carlo package {\sc emcee} for this~\cite{ForemanMackey:2012ig}. We perform the analysis using seven free parameters: the three mixing angles, two mass-splittings, the CP-violating phase $\delta$, and a constant matter density $\rho$. As the only oscillation parameter that is not reasonably well-known is $\delta$, we perform the study for four assumed physical values: $0$, $\pi/2$, $-\pi/2$, and $\pi$. Additionally, for each value of $\delta$, we perform two analyses: one in which $\rho$ is a free parameter, and one in which it is constrained by a Gaussian prior to be $(2.845 \pm 0.028)$ g/cm$^3$, corresponding to $1\%$ uncertainty. Gaussian priors are also included on $\Delta m_{21}^2 = (7.58 \pm 0.21)\times 10^{-5}$ eV$^2$ and $\sin^2\theta_{12} = 0.311 \pm 0.017$, as DUNE is not sensitive to these parameters.

\begin{figure}[!htbp]
\includegraphics[width=\linewidth]{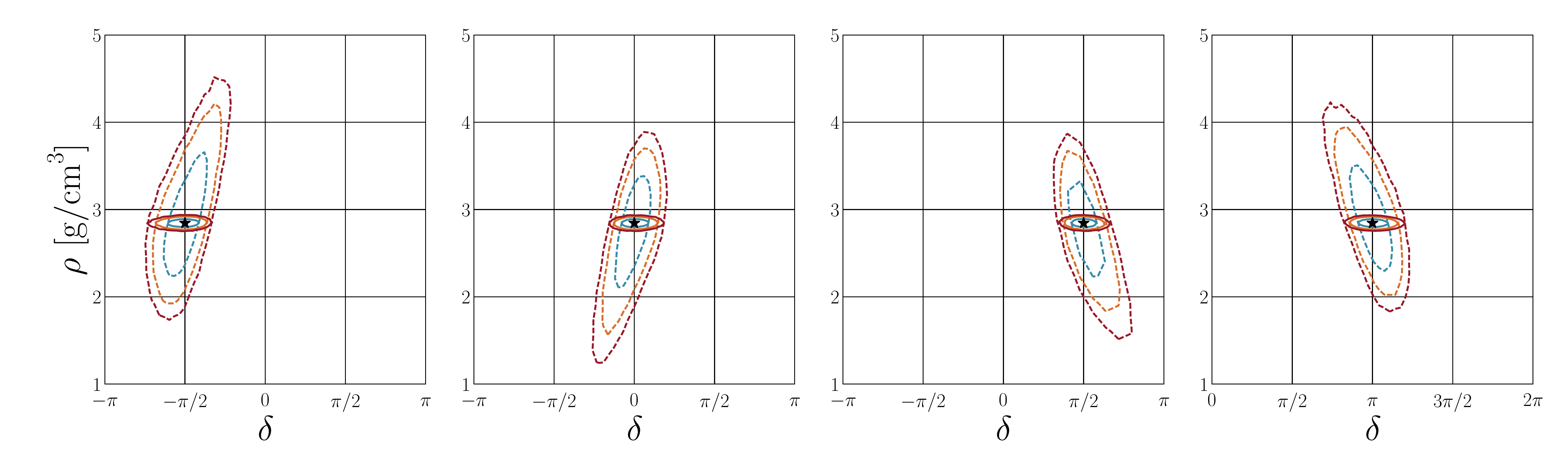}
\caption{Expected measurement sensitivity of DUNE with 300 kt-MW-yr of exposure to the parameters $\delta$ and $\rho$, where $\rho$ is assumed to be constant. Four different physical values of $\delta$ are assumed, from left to right: $-\pi/2$, $0$, $\pi/2$, and $\pi$. Stars denote the assumed physical values of $\delta$ and $\rho$ in each panel. Contours correspond to $68.3\%$ (blue), $95\%$ (orange) and $99\%$ (red) credibility regions, where all other parameters have been marginalized. Solid lines correspond to analysis including the Gaussian prior $\rho = (2.845 \pm 0.028)$ g/cm$^3$, and dotted lines correspond to analysis with $\rho$ free.}
\label{fig:Measurement}
\end{figure}

The resulting measurement capability, after marginalizing over all parameters except for $\delta$ and $\rho$ is shown in Fig.~\ref{fig:Measurement}.
Here, we have restricted ourselves to only see the $\delta$-$\rho$ plane, as there is no interesting impact on changing the assumptions on $\rho$ for any other parameter. We highlight two features here. First, the DUNE sensitivity to $\rho$, even at the $1\sigma$ level, is approximately a $25\%$ change, relative to its physical value. This is significantly larger than the $1\%$ change discussed cf Fig.~\ref{fig:ShapeNorm}, which itself dominated effects due to changing the matter density shape. Additionally, the effect on the DUNE measurement of $\delta$ is marginal: the confidence interval widens slightly for the $3\sigma$ region, however the difference between completely fixing $\rho$ to its true value and allowing it to have a $1\%$ prior is minimal.

\bibliographystyle{apsrev-title}
\bibliography{MatterBib}{}

\end{document}